# Room Temperature InP DFB Laser Array Directly Grown on (001) Silicon

**Zhechao Wang[1,2,§], Bin Tian[1,2,§], Marianna Pantouvaki[3], Weiming Guo[3], Philippe Absil[3], Joris Van Campenhout[3], Clement Merckling[3], Dries Van Thourhout[1,2,*]**

[1] INTEC Department, Ghent University, Sint-Pietersnieuwstraat 41, Ghent 9000, Belgium

[2] Center for Nano- and Biophotonics (NB-Photonics), Ghent University, Sint-Pietersnieuwstraat 41, Ghent 9000, Belgium

[3] IMEC, Kapeldreef 75, 3001 Heverlee, Belgium

* Address: Gent University, Dept. of Information Technology (INTEC), Sint-Pietersnieuwstraat 41, 9000 Gent, Belgium
  Phone: +32-9-264 3438
  Fax: +32-9-264 3593
  E-mail: Dries.Vanthourhout@intec.ugent.be

[§] These authors contributed equally to this work.

**Fully exploiting the silicon photonics platform requires a fundamentally new approach to realize high-performance laser sources that can be integrated directly using wafer-scale fabrication methods. Direct band gap III-V semiconductors allow efficient light generation but the large mismatch in lattice constant, thermal expansion and crystal polarity makes their epitaxial growth directly on silicon extremely complex. Here, using a selective area growth technique in confined regions, we surpass this fundamental limit and demonstrate an optically pumped InP-based distributed feedback (DFB) laser array grown on (001)-Silicon operating at room temperature and suitable for wavelength-division-multiplexing applications. The novel epitaxial technology suppresses threading dislocations and anti-phase boundaries to a less than 20nm thick layer not affecting the device performance. Using an in-plane laser cavity defined by standard top-down lithographic patterning together with a high yield and high uniformity provides scalability and a straightforward path towards cost-effective co-integration with photonic circuits and III-V FINFET logic.**

The potential of leveraging well-established and high yield manufacturing processes developed initially by the electronics industry has been the main driver fueling the massive research in silicon photonics over the last decade[1-6]. From the start of its development though the lack of efficient optical amplifiers and laser sources monolithically integrated with the silicon platform inhibited the widespread adoption in high-volume applications. Solutions relying on flip-chipping prefabricated laser diodes [7,8] or bonding III-V epitaxial material [9-11] are now being deployed in commercially available optical interconnects but are less compatible with standard high-volume and low cost manufacturing processes. Approaches focusing on the engineering of group IV materials have achieved optical gain but still require extensive work to reach room temperature lasing at reasonable efficiency[12-14]. Therefore, the monolithic integration of direct bandgap III-V semiconductors, well known to be efficient light emitters, with the silicon photonics platform is heavily investigated. However, considerable hurdles need to be overcome. When directly growing III-V semiconductors on silicon substrates, the large lattice mismatch ($\varepsilon_{InP/Si}$ = 8.06 %), the difference in thermal expansion and the different polarity of the materials result in large densities of crystalline defects including misfit and threading dislocations, twins, stacking faults and anti-phase boundaries, strongly degrading the performance and reducing the lifetime of any device fabricated in the as-grown layers[15]. Several routes to overcome these issues have been proposed. GaP-related materials can be grown on exact (001) silicon substrates with a small lattice mismatch and pulsed laser oscillation around 980 nm up to 120 K[16] has been achieved but shifting the laser wavelength towards the telecommunication bands remains challenging. Also GaSb, while being strongly lattice-mismatched with silicon, can be integrated on silicon without vertically penetrating threading or screw dislocations, and room temperature laser operation has been demonstrated[17]. The relatively thick buffer layer however hinders co-integration with electronic or photonic integrated circuits. Good thermal stability and high output power was shown for InAs quantum dot lasers embedded in a thick GaAs buffer[18]. But the few degrees miscut silicon substrate together with the thick III-V buffers needed to accommodate the lattice mismatch are not compatible with standard CMOS processes. Finally, also the growth of III-V nanowires on silicon is heavily studied[19,20], but it is challenging to integrate these together with low loss optical waveguide circuits.

Recently however the field of III-V epitaxy on silicon received a boost by the renewed interest of the electronics industry in using high mobility compound semiconductors in next generation CMOS[21,22]. Low defect density growth of GaAs[23], InP[24,25] and InGaAs[26] compounds using selective area growth on pre-patterned (001)-silicon was achieved, leading to the demonstration of the world's first III-V FinFET devices grown on a 300 mm substrate. Here we leverage this process to demonstrate room temperature laser operation of wafer scale integrated InP lasers directly grown on standard (001)-silicon substrates. Starting from millimeters long InP-waveguides with high optical quality grown selectively onto a silicon substrate and using standard top-down integration processes, including the definition of gratings on top of these waveguides, we fabricated Distributed Feedback Lasers (DFB) exhibiting robust single mode operation. We also demonstrated an array of DFB-lasers with well controlled emission wavelengths, illustrating the uniformity of the III-V material and devices and showing its promise for high capacity Wavelength Division Multiplexing (WDM) applications. The in-plane configuration and the use of a selective area growth technique directly on standard (001)-silicon substrates, facilitates future integration of this DFB-laser array with both photonic and electronic circuits.

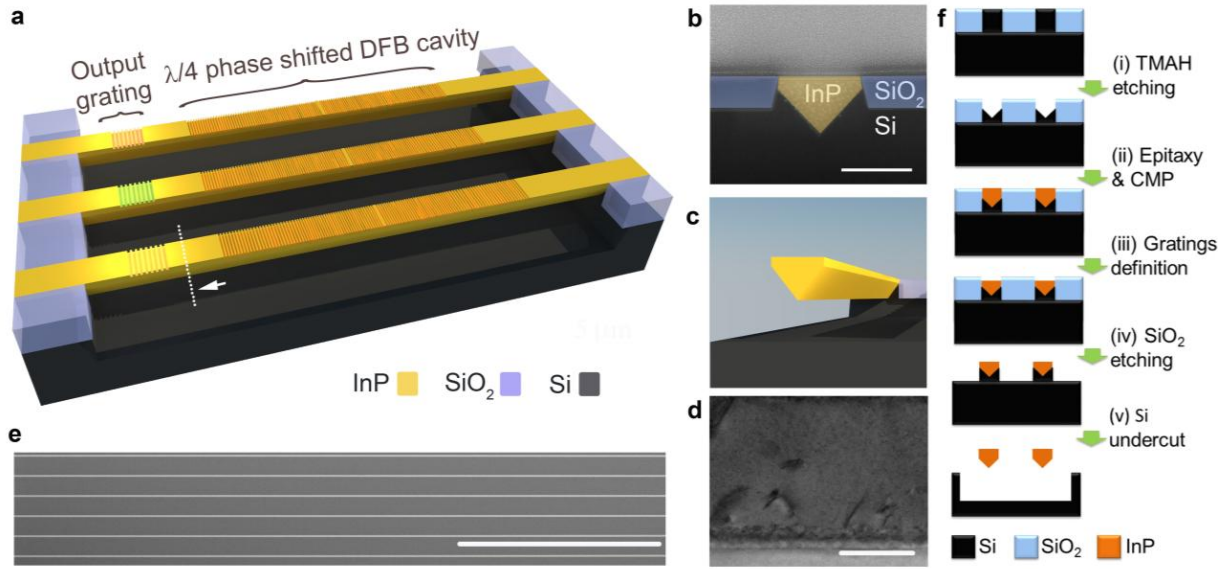

**Figure 1 Monolithic integration of InP lasers on silicon. a**, Schematic plot of the monolithically integrated InP DFB lasers on silicon. The laser cavities and the output gratings are labeled. Differently colored output gratings illustrate the tunability of the lasing wavelength. **b**, False colored SEM cross-section view of an InP-on-silicon waveguide. The scale bar is 500 nm. **c,** Schematic cross-section plot of the diamond-shaped waveguide, the position where this cross-section is taken is marked by the dashed line in Fig. 1a. **d**, TEM image of a specimen prepared parallel to the InP-on-silicon waveguide. The scale bar is 200 nm long. **e**, SEM top view of an array of InP-on-silicon waveguides. f, Integration flow used. Cross-sections orthogonal to the laser axis following each separate process step are shown.

## Results

**Fabrication.** Fig. 1a schematically depicts the fabricated DFB-laser array. Each laser consists of a high-quality InP-waveguide, on top of which 1$^{st}$ order gratings are defined with a λ/4 phase shift section inserted in the center. To facilitate characterization of the devices, 2$^{nd}$ order gratings coupling the light vertically out of the laser bar are defined 30 µm away from the DFB cavities. Since the InP is directly grown on silicon and its bandgap energy (1.35 eV @ 300K) is higher than that of silicon (1.12 eV @ 300 K), all light generated in the III-V material would directly be absorbed in the substrate when pumping the as-grown structures. Therefore the silicon substrate beneath the InP-based laser device is intentionally etched away. The suspended laser cavity is supported by two silicon pedestals outside the laser cavity. Fig. 1b shows the scanning electron microscope (SEM) cross-section view of a single waveguide before the silicon substrate undercut. The InP is 500 nm wide at the top, and the thickness of the SiO$_2$ layer next to the waveguide is 250 nm. Note the particular diamond-shape of the InP-waveguide, which it inherits from the epitaxial process employed[27]. Fig. 1f (step i to ii) briefly summarizes the epitaxy process flow. The starting substrate is a 300mm (001) silicon wafer on which 500nm wide ridges planarized with silicon oxide are defined in a standard shallow-trench-isolation (STI) process. Then, the silicon ridges are selectively etched away using a Tetramethylammonium hydroxide (TMAH) solution (5% @ 80°C), forming a V-groove composed of two flat <111> planes at the bottom of the resulting trench[27]. The subsequent InP selective area growth inside this trench is carried out in a metal organic chemical vapor phase epitaxy (MOPVE) reactor and consists of a low temperature nucleation step followed by a higher temperature growth as described in the Methods. This second step is continued until the InP sticks out above the SiO$_2$-mask after which the top surface is planarized by a chemical-mechanical-polishing process (CMP). A top view SEM image of the grown waveguide array taken after this step is shown in Fig. 1e. These uniform waveguides can be several millimeters long.

Given the 8% lattice mismatch as well as the 84% difference in thermal expansion coefficient, growing InP directly on silicon typically results in a large density of misfit and threading dislocations, detrimental for optoelectronic devices[15]. A common approach to accommodate the mismatch uses a several micrometer thick buffer layer[28-30]. However such a thick buffer layer inhibits integration with silicon electronic or photonic devices fabricated on the same wafer. Therefore, in our approach, we limit the buffer layer thickness to approximately 20nm. Nevertheless, the transmission electron microscope (TEM) image of Fig. 1d shows that the InP waveguide grown by the selective area growth approach proposed here is of high crystalline quality despite the very large lattice mismatch. The cross-section TEM lamella is prepared along the waveguide axis (parallel view) giving a good overview of the epitaxial quality along the length of the waveguide. Except for the first 20nm thick dark layer at the InP-Si interface and some stacking faults along the trench originating from the complex growth process, hardly any dislocations can be found in the material. The bottom defective layer is formed during the early stage of the epitaxy process[25]. The well-controlled low temperature nucleation process accommodates the entire lattice mismatch within a few tens of nanometers, allowing for fully relaxed and threading dislocation-free InP growth in the next step. Careful analysis of the optical modes supported by this waveguide (see Complementary Information) shows that the fundamental optical mode exhibits only limited overlap with this defective layer and is mainly located in the high quality upper region of the ridge. The damage induced by the CMP process at the top of the ridge in principle could be removed by a soft wet etching process[31], providing room for future performance improvement. Occasionally we find extended planar defects, e.g. micro-twins and stacking faults (SFs) in the bulk region of the waveguide. They are believed to be formed when the growth fronts of two spatially separated nucleation sites coalesce. PL measurements (see Complementary Information) show no emission from sub-bandgap states typically associated with defects though, proving the high quality of the material.

Besides the lattice mismatch, the polarity mismatch between InP and silicon could lead to additional defects through the formation of anti-phase boundaries (APBs), which can extend vertically towards the top surface[32,33]. By initiating the growth from the flat <111> Silicon planes at the bottom of the trench the formation of such anti phase-domains is prevented and the TEM images showing that the grown InP is indeed APB-free[27].

After the hetero-epitaxial growth, the 300 mm silicon wafer was diced into small dies for further processing (steps iii to v in Fig. 1f). The fabrication starts with the definition of the gratings in the InP-waveguides by electron beam lithography (EBL) and inductively coupled plasma (ICP) dry etching. Then, a carefully optimized isotropic reactive ion etching (RIE) process was used to remove the silicon substrate below the lasers (see Methods). By avoiding wet etching processes, the 200 µm long, 500 nm wide InP waveguides show no collapse to the substrate. While in this work EBL is used for the definition of the gratings, the critical dimensions are easily attainable using deep UV-lithography widely used in the CMOS-industry, inhibiting in no way scaling up the fabrication to high volumes.

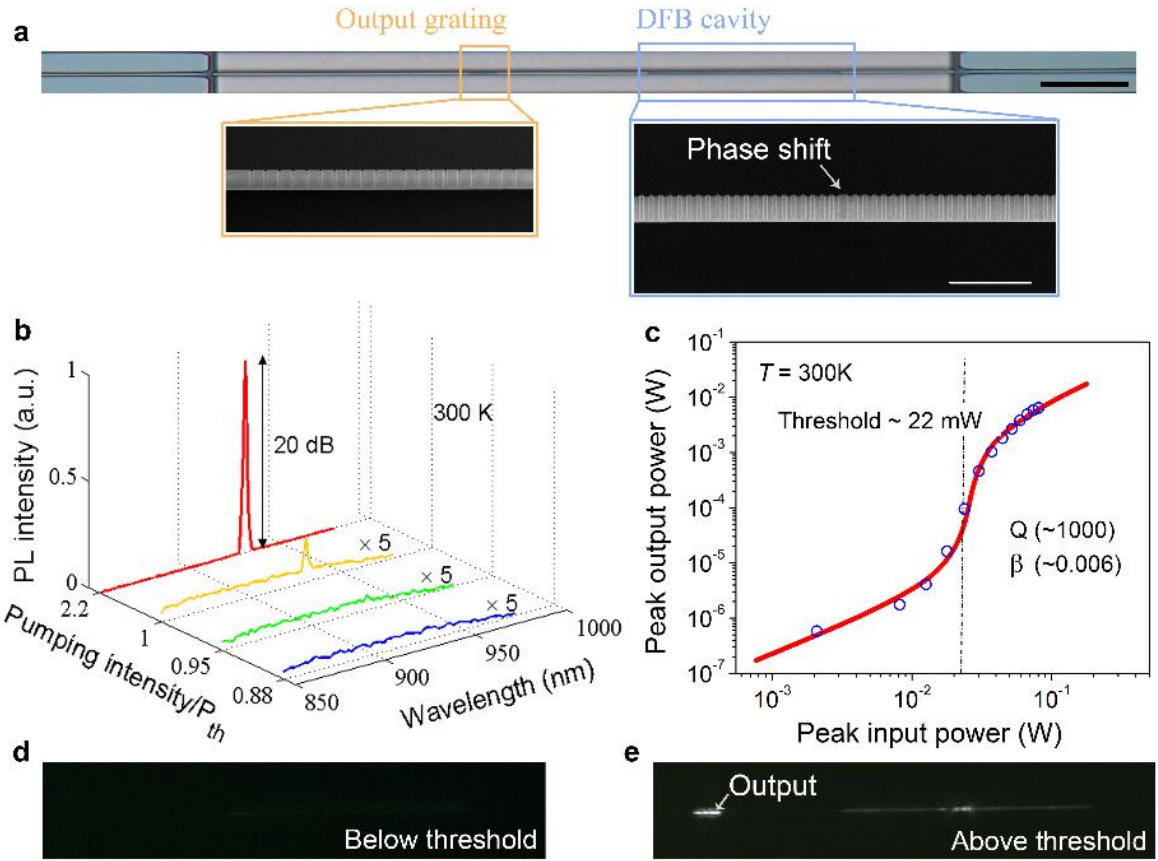

**Figure 2 Laser operation single monolithically integrated laser. a**, Microscope image of a suspended DFB laser grown on silicon substrate. The DFB cavity and the output gratings are outlined by rectangles. The scale bar is 20 μm long. Insert: SEM images of the DFB cavity and the output grating. The scale bar is 2 μm long. **b**, Room temperature laser emission spectra measured at different pump intensities. The first three emission spectra have been magnified ×5 for better visibility. A 20 dB background suppression ratio is achieved when the pumping intensity reaches 3 times the threshold level. **c**, Light in–Light out curve of the measured DFB laser. Dots are the measured data, while the curve is the rate equation fit. A spontaneous emission factor β~0.006 is extracted. **d,e** Camera recorded PL images of the laser cavity below and above threshold, respectively.

**A monolithically integrated InP DFB Laser on silicon.** While the diamond-shaped InP waveguide (Fig. 1c) supports multiple transverse modes, the large effective index difference between the fundamental mode and the higher order modes makes it possible to design a DFB cavity supporting single mode lasing (see detailed mode analysis in the Complementary Information). The laser structures were simulated using a three-dimensional finite difference time domain (3D-FDTD) method. Fig. 3c shows the calculated stop band for a grating etched 60nm deep in the InP waveguide as function of its period. The two grey lines denote the lower and upper band edge respectively. The dashed line in the center of the band indicates the expected DFB laser operating wavelength, assuming an exact λ/4 phase shift is introduced[34]. Based on these simulations we fabricated λ/4 phase shifted DFB lasers with a grating period of 163 nm and etch depth of 60nm. The total cavity length is 45 μm (see microscope and SEM images in Fig. 2a). The grating coupling strength $\kappa$L is calculated to be 2.87. The lasers are characterized using a micro-photoluminescence (μ-PL) setup (see Methods). A 6 μm wide uniform pumping area defined using a spatial filter fully covers the DFB cavity. Fig. 2b shows the PL-spectra of the device optically pumped at room temperature using a Nd:YAG nanosecond pulsed laser. When the pump intensity is low, a broad spontaneous emission spectrum is measured. By increasing the pump intensity, an optical mode starts to build up and the full width at half maximum (FWHM) of the resonant peak narrows down. This mode, centered at 930.5nm ultimately reaches about 20 dB above the background, a clear evidence of laser operation. Fig. 2c plots the peak output power as a function of the peak pump power on a logarithmic scale (L-L curve). To estimate the peak power, we assumed the power level to be constant during the 7 ns pulse, and the measured output power has been carefully calibrated to include the PL setup system loss, the coupling efficiency of the 2nd order grating and the transmission through the waveguide from the DFB-laser to the grating (see sections IV to VI in Complimentary Information). The clear discontinuity in the slope of the curve is another strong signature of laser operation. The laser threshold ($P_{th}$) is determined to be 22±2.1 mW. When increasing the pump intensity to $3P_{th}$, a peak output power of 6.4 mW is measured, equivalent to an external efficiency of 6%. From a static rate equation fitting model (blue solid curve in Fig. 2c), we extract a spontaneous emission factor (β) of 0.006, and a cavity quality factor (Q) ~ 1000. We suspect that the difference with the calculated 'cold' cavity Q factor of 6000 can be explained by fabrication imperfections degrading the optical performance of the cavity. Fig. 2d and e present the PL images of the DFB laser recorded below and above threshold, respectively. While operating below threshold, the low PL emission from the DFB cavity can be seen barely. Above threshold (Fig. 2e), a bright spot in the center of the DFB cavity appears. The spatially concentrated mode profile is indicative of the dominance of the defect mode oscillation in this strongly coupled λ/4 shifted DFB laser (*κ*L =2.87). At the left side of the picture also the output grating, which was invisible in the below threshold picture, is now very bright, showing orders of magnitude higher output power. The 1.6 nm FWHM measured for the lasing peak shown in Fig. 2b is a typical value for all devices characterized. It is well known that wavelength chirp will appear when a semiconductor laser is modulated[35]. The varying carrier density changes the refractive index and the optical length of the laser cavity, resulting in an emission wavelength that varies throughout the duration of the optical pumping. In the current case, since the silicon detector integrates the received signal over the whole pulse duration, the measured spectrum is therefore broadened by the wavelength chirp (see the dynamic rate equation modeling in the Supplementary Information). The calculated wavelength chirp is in the order of 1 nm, which agrees with the measurement results.

It is worth mentioning that the surface damage induced by the soft dry silicon under-etching process is measured to be small. Together with the orders of magnitude lower surface recombination velocity of InP compared to most other III-V materials[36], the demonstrated laser cavity is very robust against non-radiative surface recombination, and requires no extra passivation layers to achieve efficient laser oscillation. To the best of our knowledge, this is the first time an in-plane InP laser is directly grown on silicon without any buffer layer, which is essential in facilitating coupling of light towards photonics circuits defined on the same wafer because it allows for perfect vertical alignment of the different waveguide layers.

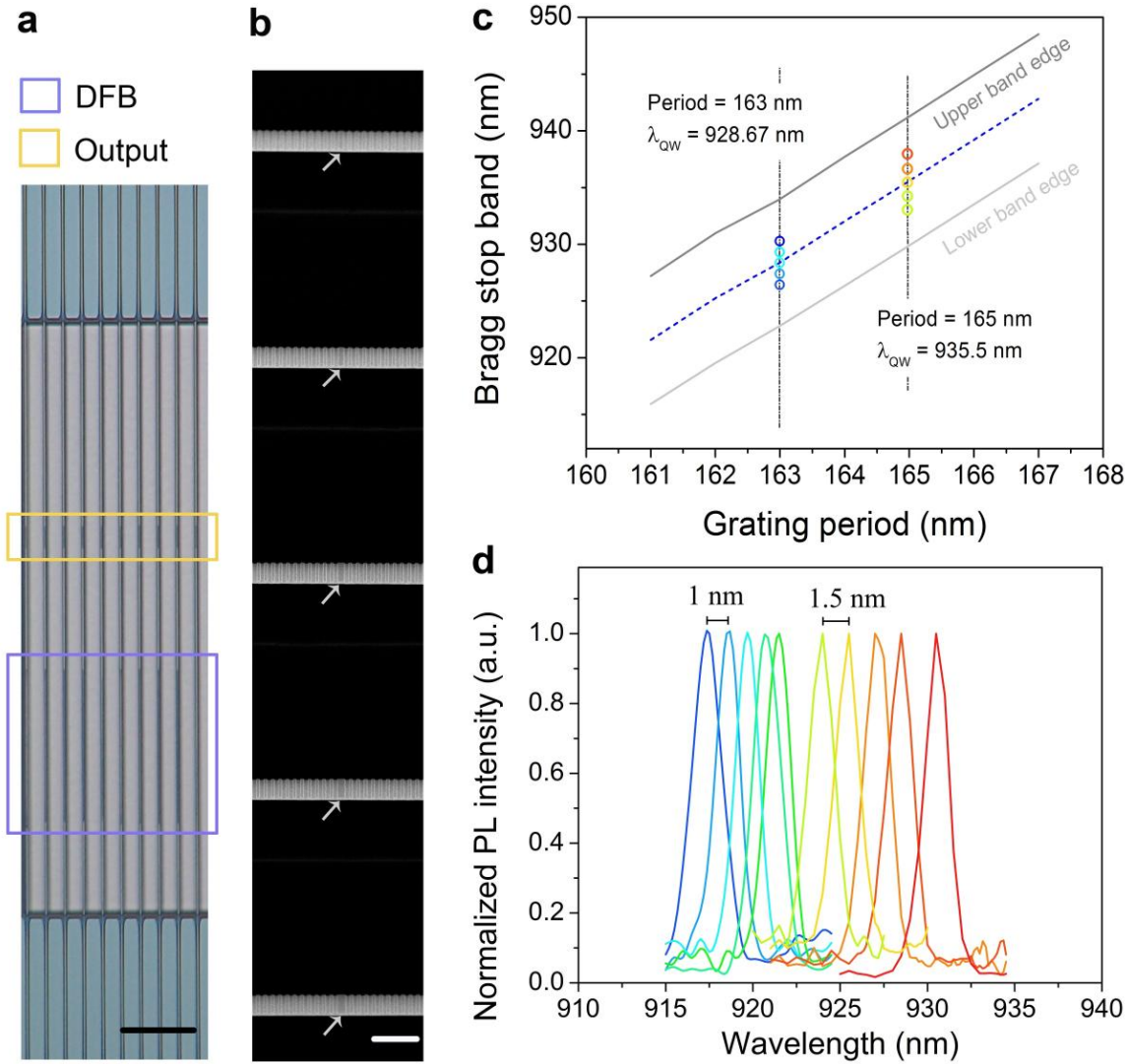

**Figure 3 DFB laser array with lithographically controlled laser emission wavelengths. a**, Optical microscope image of the DFB laser array integrated on silicon. The DFB cavities and output gratings are labeled by rectangles. The scale bar is 20 μm long. **b**, A SEM image of 5 parallel DFB lasers with the same grating period. The length-varying phase shift sections are indicated by arrows. The scale bar is 1 μm long. **c**, Calculated Bragg stop bands of the DFB grating. The grating etch depth is fixed to be 60 nm and a 50% duty cycle is used for all simulations. The dashed line in the center of the band represents the lasing wavelength $\lambda_{QW}$ for devices with an exact λ/4 phase shift in the middle of the DFB cavity. Changing the length of the central phase shift away from the λ/4 phase condition, shifts the lasing wavelength within the stop band as schematically represented by the colored dots for the two grating periods employed (163 nm and 165 nm). **d**, Measured lasing spectra from the DFB laser array shown in Fig. 3a demonstrating the capability to control the laser wavelength through grating design.

**A monolithic DFB laser array.** The high bandwidth provided by wavelength-division multiplexing (WDM) could make future optical interconnects far superior over their electronic counterparts[6] or currently deployed single wavelength optical links. Therefore, laser arrays with good wavelength control are highly demanded. Here, we demonstrate a monolithically integrated InP DFB laser array on silicon. The most straightforward way to tune the operating wavelength of a DFB laser is to vary its grating period. However, as can be seen from Fig. 3c, the latter must be controlled very accurately to achieve narrow wavelength spacing. E.g., for a 1nm wavelength step, the grating period variation will be less than half a nanometer, requiring advanced lithography techniques. To relax the requirements on the lithography tool, instead of altering the grating period, the length of the phase shift section has been altered, in steps of 20nm away from the exact λ/4 phase shift length. This small perturbation will cause a shift of the resonant wavelength within the range of the Bragg stop band (see the dots in Fig. 3c). Fig. 3a shows an optical microscope image of a silicon integrated InP DFB laser array consisting of ten lasers divided into two groups, with grating periods 163 nm and 165 nm respectively. The superimposed laser emission spectra are plotted in Fig. 3b, normalized in peak power. The lasers with 163 nm and 165 nm grating period show a uniform wavelength spacing of 1 nm and 1.5nm respectively. The small deviation of the lasing wavelength from the theoretical design (see Fig. 3c) can be attributed to fabrication errors in the grating duty cycle or other geometrical parameters. To the best of our knowledge, this is the first time that a precisely controlled in-plane laser array is monolithic integrated on silicon, which is very promising towards realizing high bandwidth communication systems requiring WDM.

The high yield – at least 98% of over 200 characterized devices showed laser operation – proves the quality of the material and the stability of the laser operation. To illustrate this further, a laser array consisting of 5 devices with constant grating parameters (165 nm period, 60 nm etching depth, exact λ/4 phase shift) was fabricated (Fig. 4a). All 5 devices show laser operation and the extracted laser operating wavelengths are plotted in Fig. 4b. The small deviation in laser wavelength between the different devices is believed to originate from the electron beam lithography tool.

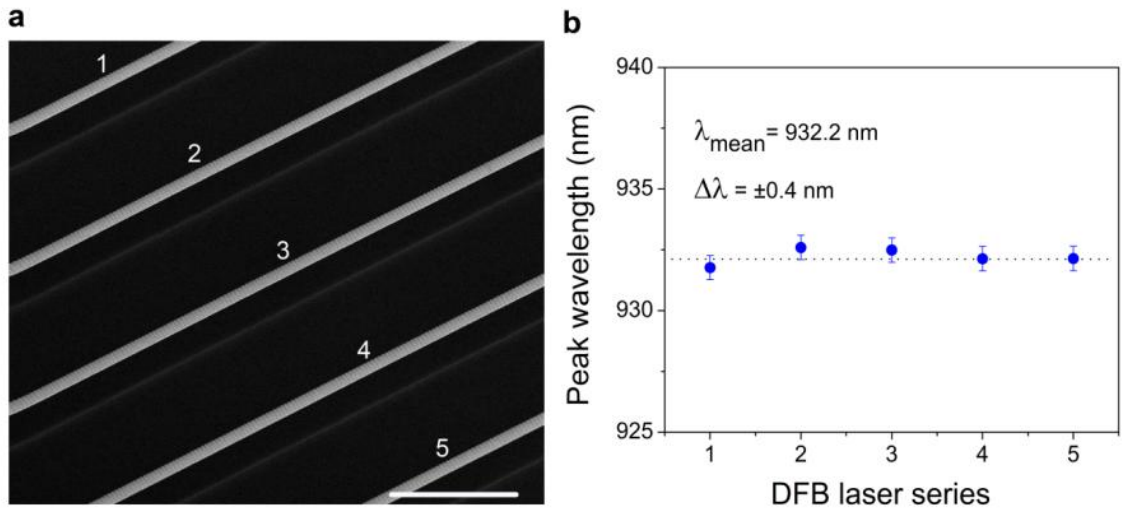

**Figure 4 High yield and scalability of the silicon integrated laser array. a**, Tilted SEM image of a laser array integrated on silicon. The cavity design is the same for all the five DFB lasers. The scale bar is 5 μm long. **b**, The peak lasing wavelengths extracted from the measured lasing spectra of the 5 identical DFB lasers (see Fig. 4a) under the same pump intensity. The tight uniformity of the emission wavelength illustrates the high yield and the scalability of the proposed monolithically integrated InP lasers on silicon.

## Discussion

We presented a monolithic integration platform promising for large scale, high volume integration of III-V lasers on silicon. Since the employed selective area epitaxy method does not limit the length of the active area, traditional and well understood laser configurations such as Fabry-Perot, distributed Bragg reflector (DBR), and DFB type cavities can be easily adopted. This is a tremendous step forward compared to for example nanowire based devices. Both the limited cavity volume and the complex integration scheme make the latter less attractive for practical applications[37]. Compared to methods relying on blanket deposition of materials the selective growth process allows co-integration with other electronic or photonic integrated devices preprocessed on the same wafer. Furthermore, since no thick buffer layer is required and the light is emitted in the plane of the wafer, one could envisage butt-coupling the proposed lasers with optical waveguides defined at the same level.

A next critical step towards a practically useful laser is demonstrating electrical injection and shifting the emission wavelength allowing to use silicon waveguide structures. Again, the in-plane laser configuration employed makes it straightforward to adopt well studied electrical injection schemes. The missing element thereby is the availability of narrow bandgap ternary or quaternary compound semiconductors that allow confining the injected carriers and red-shift the emission wavelength. This roadblock can be resolved by using the InP as a buffer on which hetero-structures, quantum wells or even quantum dots can be grown, making this virtual InP substrate a powerful platform for various applications. Growth of InGaAs on such virtual InP substrates has already been demonstrated but the optical quality of the material remains to be proven[26]. In addition the use of hetero-structures will significantly reduce carrier diffusion into the defective InP/Si interface, suppressing non-radiative recombination and improving the pumping efficiency further.

Finally, the employed silicon substrate undercut process was mainly needed to avoid high leakage loss towards the substrate. Growing the III-Vs on silicon-on-insulator (SOI) wafers, where the silicon oxide buffer underneath can effectively isolate the optical mode from the substrate could solve this issue. In addition, the direct contact of the III-V lasers with SOI will improve the thermal dissipation of the device, boosting its performance.

In summary, we have presented the first highly scalable monolithic solution to the long-awaited missing piece of silicon photonics: integrated laser sources on silicon. The in-plane configuration and the use of a selective area growth technique directly on 300 mm (001) silicon substrates in combination with a top-down integration scheme provide a route towards the integration of dense arrays of III-V laser sources with silicon photonic circuits. A wide range of applications from various fields requiring low-cost on-chip laser sources could benefit from this. In particular, for on-chip optical interconnects, the demonstrated monolithic laser array, together with the WDM technology, may finally pave the way to tera-scale computing.

## Methods

**InP epitaxial growth on pre-patterned silicon wafer.**
Standard $SiO_2$ shallow-trench-isolation (STI) patterning was applied on 300 mm on-axis Si (001) substrates to realize 250 nm thick ridges buried in the $SiO_2$ buffer. The subsequent selective etching of Si by Tetramethylammonium hydroxide (TMAH) solution results in trenches with widths ranging from 40 nm to 500 nm. The average active area exposed to the III-V layers deposition was kept constant at 10% of the total Si wafer surface. The trenches are aligned along the [110] and orthogonal [-110] directions. The III-V heteroepitaxy has been performed in a 300 mm production AIXTRON Crius Metal Organic Vapor Phase Epitaxy reactor, equipped with a vertical showerhead injector. The group-III precursors in the tool are trimethylindium (TMIn), trimethylgallium (TMGa) and trimethylaluminium (TMAl), and the group-V sources are tertiarybutylarsine (TBAs) and tertiarybutylphosphine (TBP). During the growth, the reactor pressure can vary from 50 mbar to 500 mbar with a $H_2$ total flow of 48 slm as carrier gas. First, the native oxide from the trench bottom surface is thermally desorbed by a high temperature bake above 800°C in $H_2$ at 50 mbar. Then, the substrate temperature was cooled down to a low temperature (< 380°C) to grow the nucleation layer. Below 550°C, the surface is exposed to Arsenic at high pressure to form one monolayer of As-terminated surface which will promote the InP wetting at low temperatures. After the nucleation step, the temperature was ramped up to 550°C to obtain a high crystalline quality InP layer.

**InP laser array process flow.** To fabricate the suspended DFB laser array on silicon, electron beam lithography (EBL) was first utilized to define the DFB gratings together with the λ/4 phase shift on the InP waveguide. The grating pattern was then transferred by reactive ion etching (RIE) down to the 20 nm thick $SiO_2$ hard mask that was deposited by plasma enhanced chemical vapor deposition (PECVD) prior the EBL process. The subsequent inductively coupled plasma (ICP) etching process further transferred the grating pattern to the InP waveguide. As the next step, after dipping the sample in a 2% HF solution for 20 minutes to remove the STI oxide mask and expose the silicon substrate, another 200 nm thick $SiO_2$ hard mask was deposited on the sample by PECVD. Optical lithography was then employed to define 50 µm wide resist strips at both ends of the DFB cavity, without overlapping on pre-defined gratings. A RIE $SiO_2$ etch process was used to transfer the resist pattern into the 200 nm $SiO_2$ hard mask, and after removing the resist residuals, the exposed silicon substrate under the DFB laser was undercut of 3µm by an isotropic RIE etching process. No collapse or breakdown of the suspended InP cavity was observed after the RIE silicon etching process.

**Laser characterization.** The laser measurement was carried out at room temperature, without the use of any cryostat or substrate temperature controller. To achieve laser operation, 7 ns pump pulses from a Nd:YAG nanosecond pulsed laser (Ekspla, 532 nm, repetition rate 938 Hz) were delivered to the sample surface by a ×50, 0.65 numerical aperture (NA) objective (Mitutoyo NIR HR). The use of optical attenuators together with a half-wave plate and an optical polarizer allow a continuous tuning of the pump power intensity. In addition, the optical setup is designed to deliver a rectangle uniform pump pattern that fully covers the whole DFB cavity without any overlay on the adjacent devices. The DFB laser emission was collected by the same objective used for pump delivery and detected through a ¼ m monochromator (MS257, Newport) by a thermoelectric (TE)-cooled silicon detector. Filters were used to block the pumping light from reaching the detectors. The detector signal was locked by a lock-in amplifier to improve the signal-to-noise ratio.

**Acknowledgements:**

This work is supported by the European Commission through the ERC project: ULPPIC (Ultra Low Power Photonic IC) and imec's industry-affiliation program on Optical I/O. The authors thank R. Baets, G. Roelkens and N. Le Thomas for fruitful discussions.

**Supplementary Information**

**I.   Optical mode analysis**

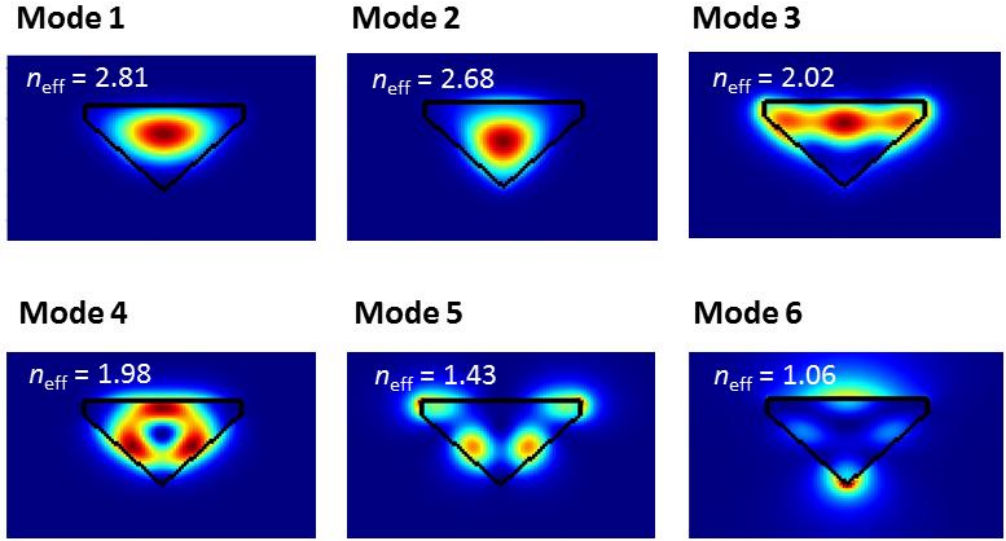

**Figure S1. Simulated optical mode profiles (electric field intensity) that are supported by the diamond-shaped InP waveguide.**

By using a full-vectorial finite difference (FD) mode solver, the mode profiles (electrical field intensity) of the supported modes of the diamond-shaped InP waveguide are calculated and plot in Fig. S1. The removal of the silicon substrate underneath is essential to obtain guided modes with negligible leakage loss. Due to the asymmetrical waveguide shapes, the modes are not purely transverse-electric (TE) polarized or transverse-magnetic (TM) polarized, but are hybrid modes. The fundamental mode is more TE-like, while the first order mode is more TM-like. In Fig. S1, the calculated effective refractive indices of the optical modes are also provided. Although the waveguide supports multiple transverse modes, their effective refractive index difference is large enough to well separate their Bragg grating stop bands, ensuring that only the fundamental mode reflection peak overlaps with the material gain spectrum. Therefore single transverse-mode lasing can be achieved in the proposed DFB cavity.

In addition, as presented in the manuscript, due to the lattice mismatch, the InP/Si interface is highly defective. By carefully inspecting the mode profile of the fundamental mode, one finds that most of the optical energy is confined in the bulky InP region, and therefore the influence of the defective layer on the modal gain is minimized.

## II.  Photoluminescence study of the InP epitaxially grown on silicon

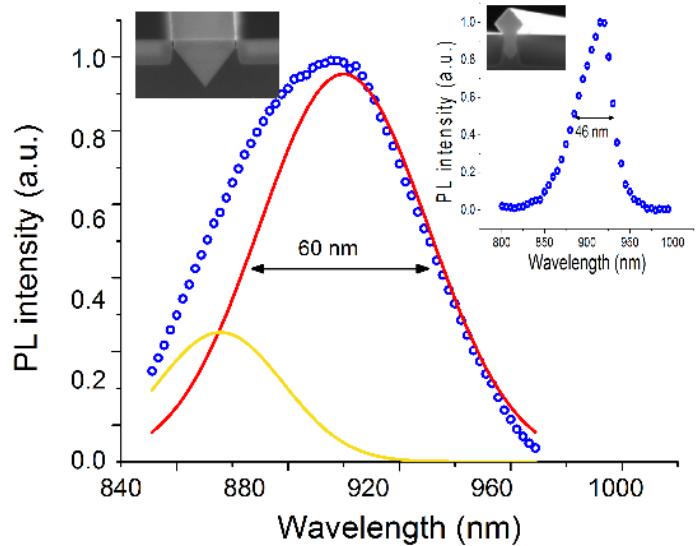

**Figure S2. Photoluminescence study of the InP grown on silicon.** Normalized Photoluminescence (PL) spectrum (blue open circles) of an array of 500 nm wide InP-on-Si waveguides after CMP. A double Gaussian curve fit has been applied to decompose the PL spectrum into two peaks (red and yellow curves). Insert: Normalized PL spectrum measured from an array of 500 nm InP-on-Si waveguides with overgrown InP on top.

While TEM analysis provides direct but scattered information on the material quality, photoluminescence (PL) characterization is a powerful tool that provides valuable information on the overall material quality. The room-temperature PL spectrum of an array of 500 nm wide InP-on-Si waveguides after the CMP-process is presented in Fig. S2. The pump source is a continuous wave (CW) 532 nm laser source. From a double Gaussian curve fitting process two curves centered at 910 nm and 870 nm, respectively are found. The relative wide linewidth of the band edge emission peak (910 nm) is believed to originate mainly from the thin defective interface between Si and InP. To support this, the insert shows the PL spectrum measured from an array of InP-on-Si waveguides with thick overgrown InP on top (see the insert SEM image). Given the strong absorption of the 532 nm pumping light in InP, the pumping efficiency of the defective layer is very weak in this case. The narrow linewidth (46 nm) of this PL spectrum proves that, except for the defective interface, the grown InP material has superior crystalline quality.

It is interesting that both PL spectra have a shoulder on the higher energy side of the band edge emission peak. Considering that defects normally emit photons with energy lower than the material bandgap, it can be concluded that this high photon energy emission is not related to dislocations present in the material. The exact origin of this higher energy shoulder is still under investigation. However, it is found that twins and stacking faults are present in the bulk of the InP material. A possible explanation therefore is that the valence band splitting of the wurtzite-like InP crystal phase formed by the twins is responsible for this high photon energy emission peak[38,39]. From our previous work[40] and the reliable laser oscillation demonstrated in this work, we can conclude this InP crystal phase mixture does not comprise the material's quality for optoelectronic applications.

## III. Wavelength chirp - dynamic rate equation model

To gain more insight in the measurement results, in particular, the relatively large FWHM of the lasing peak, we used a dynamic rate equation model to explore the laser behavior under pulsed pumping conditions:

$$\frac{dN}{dt} = \frac{P_{pump}}{E_p V_{active}} - \frac{S_{sur} V_{sur}}{V_{active}} N - BN^2 - CN^3 - V_g GS \quad (1)$$

$$\frac{dS}{dt} = \Gamma V_g GS - \frac{S}{\tau_p} + \Gamma \beta BN^2 \quad (2)$$

where the optical gain G is defined as

$$G = \frac{g_0(N - Nt)}{1 + \varepsilon S} \quad (3)$$

The definition and where relevant the values of the parameters used are listed in the table below:

Table 1. Rate equation parameters and constants

|  | Parameters | Value (only for constants) |
|---|---|---|
| N | Carrier density |  |
| S | Photon density |  |
| G | Optical gain |  |
| t | Time step |  |
| $P_{pump}$ | Pump power |  |
| $E_p$ | Emitted photon energy | 3.75e-19 (J) |
| $V_{active}$ | Active region volume | 4.58e-18 (m³) |
| $S_{sur}$ | Active region surface area | 5.09-11 (m²) |
| $V_{sur}$ | InP surface recombination velocity | 1e4 (cm/s) |
| B | bimolecular recombination coefficient | 2e-16 (m³/s) |
| C | Auger coefficient | 8e-41 (m⁶/s) |
| $V_g$ | Group velocity | 6.82e7 (m/s) |
| Γ | Optical confinement factor | 0.85 |
| $\tau_p$ | Photon lifetime | 0.49 (ps) |
| β | Spontaneous emission factor | 0.007 |
| $g_0$ | Differential gain | 3e-20 (m²) |
| ε | Gain compression factor | 1e-23 (m³) |
| $N_t$ | Transparency Carrier density | 1.23e18 (cm⁻³) |

It is widely known that modulation of a semiconductor laser induces a dynamic change of the carrier density in the laser cavity. Due to the plasma effect, this carrier density variation results in a wavelength chirp that can be calculated as:

$$\Delta \lambda = \Gamma \frac{1}{4\pi n_g} g_0 \alpha \lambda^2 (N - Nt) \quad (4)$$

where $n_g$ is the group index, which is 4.4 in the current case. A value $\alpha=4$ is chosen for the linewidth enhancement factor. Fig. S3 presents the simulated carrier density, photon density, and the corresponding wavelength chirp as a function of time, when a 7 ns pulse (FWHM) is delivered to the sample. The pulse energy is set to be about three times the threshold measured for this laser.

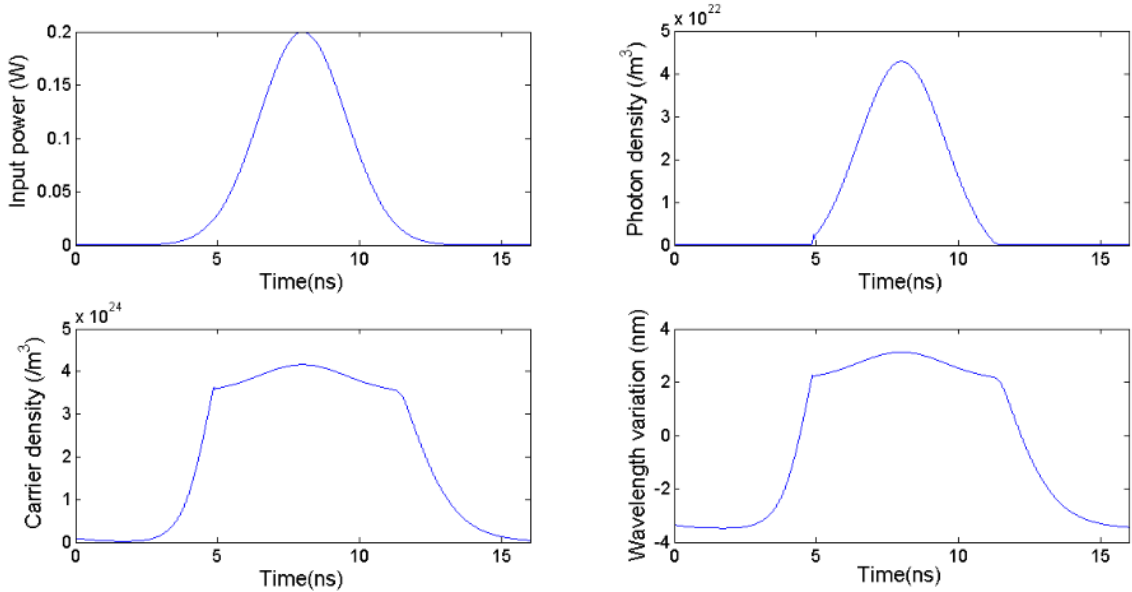

**Figure S3. Rate equation analysis of the dynamic response of the laser cavity to a single pump pulse.** The input pumping power, photon density, carrier density, and wavelength variation are plot as a function of time.

It is found that under pulsed pumping conditions, the carrier density doesn't clamp above threshold. The carrier density variation above threshold is $0.6\times10^{24}$ m$^{-3}$, and the corresponding wavelength chirp is 0.83 nm. Since the wavelength chirp is proportional to the carrier density change (see equation 4), it also explains the observed broadening of the laser peak linewidth when increasing the pump intensity further above the threshold.

## IV. Second order grating design (output coupler)

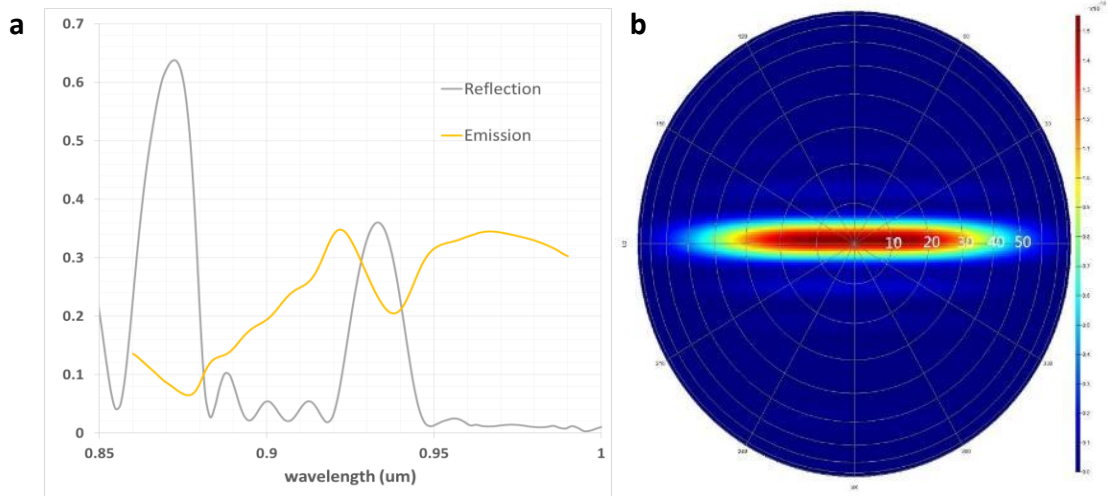

**Figure S4 Theoretical analysis of the output grating coupler. a,** Simulated out coupling and back reflection efficiencies of the grating. The gray line is the fraction of the power that is reflected back towards the cavity, while the yellow line is the out-coupling efficiency to the free space upwards. **b,** Simulated far field distribution of the output light emission.

As presented in the manuscript, a 2$^{nd}$ order grating is defined close to the DFB cavity for vertical light extraction. The grating period is set to be twice the DFB grating period, i.e. 326 nm, and the duty cycle is around 25%. The number of grating periods is chosen to be 20, as a tradeoff between the out-coupling efficiency and the back reflection. The grating etch depth is 60 nm, the same as the DFB grating. The grating was simulated using 3D FDTD. The calculated spectra for the light coupled upwards (yellow) and reflected back (grey) are plotted in Fig. S4a. This simulation shows that up to 36% of the light can be reflected at the grating of which 94.4% is coupled back into the fundamental mode. The influence of this high reflection on the laser operation is analyzed in section VI below.

To estimate the collection efficiency of the vertical laser emission, we simulated the grating emission far field (Fig. S4b). The waveguide is oriented from bottom to top in the figure. Considering the narrow waveguide dimension in the horizontal direction (500 nm), it is not surprising to find that the far field pattern expands mostly in the lateral direction. From the numerical aperture of the objective (NA=0.65), an angular aperture half-angle of 40.5 degree is derived, which effectively overlaps with the light emission far field pattern.

## V.     Waveguide absorption efficiency analysis

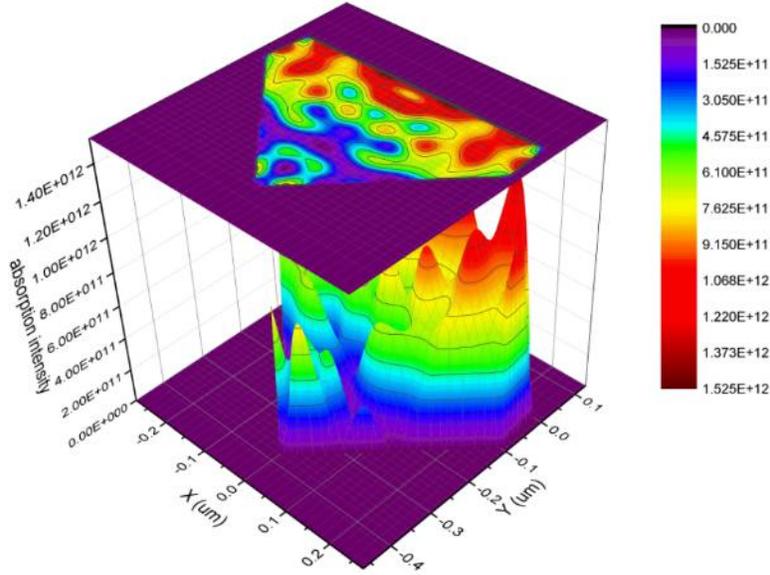

**Figure S5 Simulated light power absorption intensity distribution in the diamond-shaped InP waveguide.**

In order to calculate the fraction of the pumping power that is actually absorbed by the InP waveguide, FDTD simulations were carried out. The calculated distribution of the power absorption intensity inside the waveguide is plot in Fig. S5 assuming the sample is uniformly illuminated by a plane wave with 532 nm wavelength normally incident on the waveguide top surface. The lateral calculation window is set to be 5.5 µm wide, with periodic conditions set on the two lateral boundaries. In this way, the model mimics the case of an array of DFB lasers being uniformly pumped from the top (10% InP coverage). Fig. S5 shows a complicated power absorption distribution inside the diamond-shaped waveguide. The interference of the light reflected from the high index contrast waveguide surfaces is mainly the reason why this pattern is formed. The integration of the absorption intensity over the waveguide cross-section gives us the fraction of the light power that is absorbed by the InP waveguide, which is 7.93%.

## VI. Analysis of the effect of the output grating reflection on the DFB laser operation

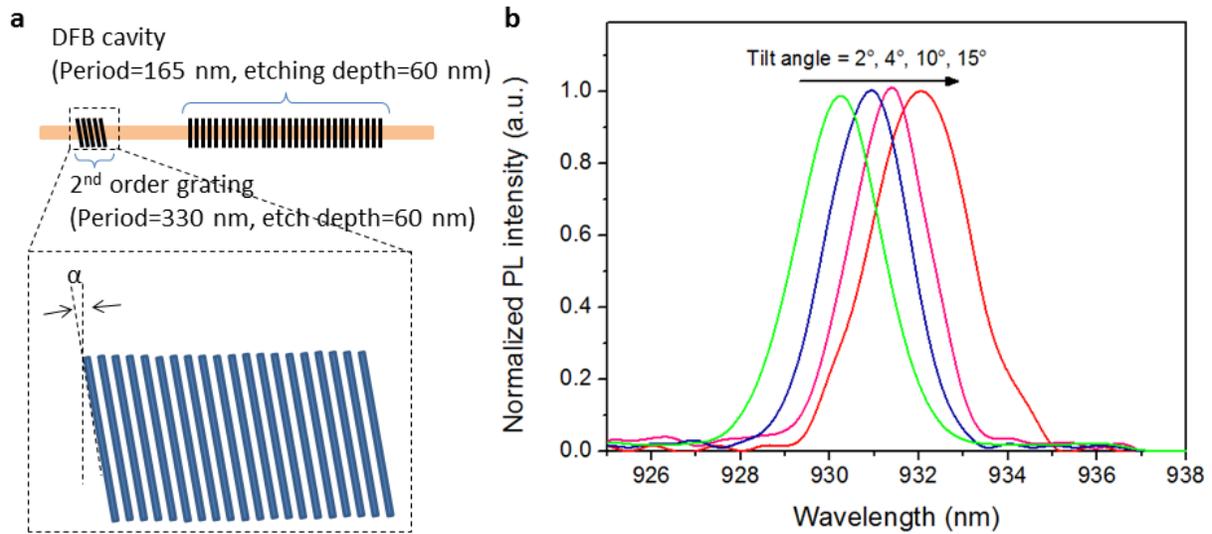

**Figure S6. The influence of the output grating reflection on the DFB laser operation. a,** A schematic plot of the DFB laser configuration. A zoom-in view of the output gratings shows that the gratings are tilted away from the vertical direction by an angle of α. **b,** measured lasing spectra from an array of identical DFB lasers with varying output grating tilt angles (α = 2°, 4°, 10°, and 15°).

As discussed in the previous section, up to 30% of the incident light power can be reflected back by the output grating, which may have a considerable influence on the operation of the DFB laser. A straightforward solution to minimize the influence of this reflection would be placing the output grating far away from the DFB cavity. However, in that case most of the laser emission will be lost in the highly absorptive InP waveguide (without pumping) before reaching the output grating. In the laser configuration presented, the $2^{nd}$ order output grating is defined 30 μm away from the DFB cavity. To avoid excessive absorption in the waveguide section connecting the laser with the grating this section is partially pumped (about 65% of this waveguide section is pumped). We estimated that this is enough to compensate for the absorption in the unpumped part of this waveguide, leaving the overall transmission from laser to grating lossless. In order to estimate the influence of the reflection on the laser operation, we processed a sample with the $2^{nd}$ order gratings being tilted as schematically shown in Fig. S6a. The tilt angle α is swept from 2 degrees to 15 degrees while the DFB laser design is kept to be the same. The above-threshold lasing spectra measured can be found in Fig. S6b. We find that the lasing wavelength is tuned in steps of approximately 0.5 nm when the tilt angle is swept. It proves the hypothesis that the reflection changes the laser oscillation conditions. It also shows the possibility of fine tuning the laser wavelength by adjusting the external grating.